\documentclass[aps,pra,twocolumn,groupedaddress]{revtex4-1}
\usepackage[T1]{fontenc}
\usepackage{hyperref}
\input{epsf}

\usepackage{graphicx}
\usepackage{epstopdf}
\usepackage{array}
\usepackage{longtable}
\usepackage{rotating,booktabs}
\usepackage{booktabs,threeparttable}
\usepackage{bm}
\usepackage{float}

\usepackage{color}

\begin{document}

\title{Angle-dependent magic wavelengths for the $4s_{1/2}\to3d_{5/2,3/2}$ transitions of Ca$^{+}$ ions}
\author{Jun Jiang}
\email {phyjiang@yeah.net}
\author{Li Jiang}
\author{Z.~W.~Wu}
\author{Deng-Hong Zhang}
\author{Lu-You Xie}
\author{Chen-Zhong Dong}

\affiliation{Key Laboratory of Atomic and Molecular
Physics and Functional Materials of Gansu Province,
College of Physics and Electronic Engineering,
Northwest Normal University, Lanzhou 730070, P. R. China
}

\date{\today}

\begin{abstract}
The dynamic polarizabilities of the atomic states with 
angular momentum $j\textgreater \frac12$ are sensitive to the angle between the 
quantization axis $\hat{e}_z$ and the 
polarization vector $\hat{\mathbf{\epsilon}}$ owing to the contribution of 
anisotropic tensor polarizabilities. The magic wavelength, 
at which the differential Stark shift of an atomic 
transition nullifies, depends on this angle. We identified the magic wavelengths for the 
$4s_{\frac12}\to3d_{\frac32,\frac52}$ transitions of Ca$^{+}$ ions at different angles between $\hat{e}_z$ 
and $\hat{\mathbf{\epsilon}}$ in the case of linearly polarized light.
We found that the magic wavelengths near 
395.79 nm, which lie between the $4s_{\frac12}\to4p_{\frac12}$ and $4s_{\frac12}\to 4p_{\frac32}$ transition 
wavelengths, remain unsensitive to the angle, while the magic wavelengths, which are  longer than
the $3d_{\frac52}\to 4p_{\frac32}$ resonant transition 
wavelength (854.21 nm), are very sensitive to the angle.

\end{abstract}

 \maketitle

\section{INTRODUCTION}

Techniques involving laser cooling and trapping of 
neutral atoms or ions have a lot of applications in quantum information
\cite{nature-tiecke14a,pr-haffner08a,m-wilpers07a, prl-monroe95a},  
high-precision frequency and spectroscopy measurements
\cite{pr-lea07a,science-wood97a}, and optical frequency 
standards \cite{science-ludlow08a,science-rosenband08a,science-hinkley13a,
pra-huang12a,nature-takmoto05a,prl-huang16a}.
However, the laser field can cause the Stark shifts. 
The problem of eliminating the Stark shifts 
can be solved by trapping an atom or ion at magic wavelengths, 
at which the Stark shifts of both the upper and lower states are the same and the 
shifts of the transition frequency vanish \cite{ye99a, katori99b}.
Also, the systematic uncertainties of high-precision measurement can be reduced
by optical trap at the magic wavelengths \cite{prl-liu15a,arXiv-becher17a}. 
In order to theoretically determine 
the magic wavelength of an atomic 
transition, accurate dynamic polarizabilities are required for the 
relevant atomic states, which consist of isotropic 
scalar and anisotropic vector and tensor parts \cite{pr-man86a,beloy09a, arXiv-becher17a}.
The anisotropic parts result 
in a light shift, which depends on not only the angular 
momentum projection $m$ but also the angle between the 
quantization axis $\hat{e}_z$ and the electric 
polarization vector $\hat{\mathbf{\epsilon}}$ of the laser. This will make accurate 
determinations of the magic wavelengths much difficult in experiments.

Due to the structure of energy-levels is simple and the 
$3d_{\frac52}$ state has a long lifetime, 
calcium ions have been chosen as one of the candidates for 
optical frequency standard 
\cite{prl-chwalla09a,opt-kensuke12a,pra-nager00a,pra-kreuter05a,pra-barton00a,pra-arora07a,pla-cham04a,pra-kajita05a}.
In a very recent experiment with a radio-frequency Paul trap, the accuracy of $^{40}$Ca$^+$ optical
clocks has achieved a level of 3.4 $\times$ 10$^{-17}$ \cite{prl-guan2016a}. 
In this experiment, excess micromotion was
identified as the biggest factor affecting the 
accuracy of the $^{40}$Ca$^+$ clock \cite{prl-huang16a}.
If the weak micromotions of trapped ions can be handled, 
such kind of $^{40}$Ca$^+$ clocks could achieve a systematic 
fractional uncertainty of about 10$^{-18}$. 
Therefore, all-optical magic trapping of ions is 
worth of being tried for minishing substantially the micromotion-induced
shifts\cite{prl-huang16a,prl-liu15a}.

The magic wavelengths of Ca$^+$ ions have been studied extensively both in theory and experiment
\cite{pra-tang13a,pra-kaur15a,prl-liu15a}. Two magic wavelengths of the $^{40}$Ca$^{+}$ $4s\to3d_{\frac52}$
$(m=\frac12,\frac32)$ clock transitions near 395.79 nm for linearly polarized light have been measured 
with very high accuracy and they agree with all existing theoretical
results very well \cite{prl-liu15a,pra-tang13a,pra-kaur15a}. However, 
these magic wavelengths are very close to the $4s_{\frac12}\to4p_{\frac32}$
and $4s_{\frac12}\to4p_{\frac12}$ resonant transition wavelengths which span from 393.366 nm to 396.847 nm. 
Therefore, they are not good for the use of magic trapping as the near-resonance light has high photon
spontaneous scattering rates which result in a high
heating process \cite{pra-haycock97a,pra-savard97a}. 
Although many other magic wavelengths of Ca$^{+}$ ions were determined
in theory, they were considered just in the situation 
of linearly polarized light with the quantization axis perpendicular to the wave
vector ($\rm{\hat{e}_z \perp \hat{k}}$) and parallel 
to the polarization vector ($\rm{\hat{e}_z \parallel \hat{\epsilon}}$). To the best of
our knowledge, the dependence of magic wavelengths 
upon the laser polarization direction has not been studied yet. 
The group in Wuhan
Institute of Physics and Mathematics in China has tried  
to measure some other magic wavelengths with far-off resonance, but the results and
theoretical values have a big difference \cite{gao17}.

In this paper, the magic wavelengths of Ca$^+$ ions
for each of the magnetic sublevel components of the $4s_{\frac12}\to3d_{\frac32,\frac52}$
transitions are identified for linearly polarized light 
based on our previous work \cite{pra-jiang17b}. The variations of the
magic wavelengths with the applied 
laser direction are determined in detail. Finally, a brief summary is given in
Sec.~III. Atomic units, $\hbar=m=|e|=1$, are used throughout this paper unless stated otherwise.

\begin{figure}[tbh]
\vspace{0.5cm}
\label{sample1}
\setlength{\abovecaptionskip}{0pt}
\setlength{\belowcaptionskip}{0pt}
\centering{
\includegraphics[width=7.8cm]{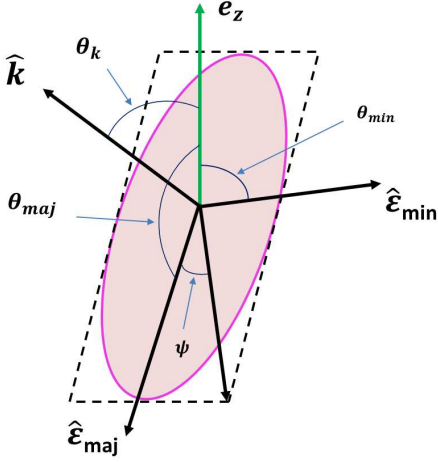}}
\vspace{-1em}
\caption{Representation of the electromagnetic plane wave geometrical parameters.
The surface represents the ellipse swept out by the electric field vector in one period.
Unit vector $\hat{\epsilon}_{\rm{maj}} (\hat{\epsilon}_{\rm{min}})$ aligns with the semi-major (-minor) axis of the ellipse.
The $\hat{e}_z$ is quantization axis which selects the direction of magnetic field in experiment.
The $\hat k$ represents the direction of wave vector.
The $\rm{\theta_k}$ is angle between the $\hat{e}_z$ and $\hat k$.
Parameters $\rm{\hat{\epsilon}_{maj}}$, $\rm{\hat{\epsilon}_{min}}$ and $\hat k$ are mutually orthogonal. 
$\rm{\theta_{maj}}$ ($\rm{\theta_{min}}$) is the angle
between the $\rm{\hat{\epsilon}_{maj}}$ ($\rm{\hat{\epsilon}_{min}}$)
and the $\hat{e}_z$.
Parameter $\psi$ is directly related to the degree of circular polarization.}
\end{figure}

\section{results and DISCUSSION}

The necessary atomic parameters of Ca$^+$ ions such as energy levels, matrix elements, 
and polarizabilities have been calculated by using
the relativistic configuration interaction plus core polarization (RCICP) approach \cite{pra-jiang16a}
in our previous work
\cite{pra-jiang17b}. These data are not repeated here for the sake of brevity.

For an arbitrarily polarized light, the dynamic polarizability of an atomic state $i$ is given by
\cite{epjd-kie13a,pr-man86a,beloy09a}
\begin{eqnarray}
\label{a}
\alpha_{i}(\omega) &=& \alpha^{S}_{i}(\omega) +
\mathcal{A} {\rm cos} \theta_{k}\frac{m_{j_i}}{2j_i} \alpha^{V}_{i}(\omega) \nonumber \\
&+& (\frac{3 {\rm cos}^{2} \theta_{p}-1}{2})\frac{3m^{2}_{j_i}-j_i(j_i+1)}{j_i(2j_i-1)} \alpha^{T}_{i}(\omega),
\end{eqnarray}
where $\alpha_i^{S}(\omega)$, $\alpha_i^{V}(\omega)$, 
$\alpha_i^{T}(\omega)$ represent the scalar, vector, and tensor polarizabilities
as given in Refs. \cite{kien13a,manakov86a,beloy09a}, 
respectively; $m_{j_i}$ is the component of total angular momentum $j_i$.
There is no tensor polarizability for the states with 
$j\leq\frac12$. $\rm{\theta_{k}}$ is the angle between the wave vector $k$ and the
quantization axis $\rm{\hat{e}_z}$, 
cos$\rm{\theta_k =\hat{k} \cdot \hat{e}_z}$. The relevant diagram is shown in Figure~1.
The $\theta_{p}$ relates to the polarization 
vector $\hat{\epsilon}$ and the $\rm{\hat{e}_z}$ axis. For a more general geometrical
interpretation of $\theta_{p}$, it is useful to 
further introduce the parameters $\theta_{maj}$, $\theta_{min}$, and $\psi $.
cos$^2\theta_p$ can be written in the form \cite{beloy09a}
\begin{eqnarray}
\label{q2}
\rm{cos}^2\theta_p = \rm{cos}^2\psi \rm{cos}^2\theta_{maj}+\rm{sin}^2\psi \rm{cos}^2\theta_{min} ,
\end{eqnarray}
where the parameter $\theta_{maj}$ ($\theta_{min}$) is the angle between the major (minor) axis of the ellipse and the $\hat{e}_z$ axis. 
From a geometrical consideration, $\theta_{k}$ and $\theta_{p}$
satisfy the relation cos$^2\theta_{k}$ + cos$^2\theta_{p} \leq 1$ \cite{epjd-kie13a,pr-man86a}.
The angle $\psi$ is directly related to the degree of polarization of the light.
$\mathcal{A}$ represents the degree of
polarization, which is give by
\begin{eqnarray}
\label{q4}
\mathcal{A}= \rm{sin}2\psi.
\end{eqnarray}
In particular, $\mathcal{A}=0$ corresponds to linear
polarization, while $\mathcal{A}=+1$ (or $-$1) corresponds to right- (or left-) circular polarization.
In experiment, however, $\mathcal{A}$ could not absolutely be equal to zero. In this case, the vector polarizability
contributes to the total dynamic polarizability. 
In order to get rid of the vector part in experiment, one can set cos$\rm{\theta_{k}}$ be
equal to zero that is $\rm{ \hat{e}_z \perp \hat k}$. 

For the case of cos$\theta_k=0$, the quantization axis $\hat{e}_z$ is
perpendicular to the wave vector $\hat k$, 
and the angle between the direction of polarization and $\hat{e}_z$ 
varies in the plane of polarization. 
When $\mathcal{A} =0$, or cos$\theta_{k}$ = 0, 
the dynamic polarizability can be easily simplified from Eq.(\ref{a}) as follows,
\begin{eqnarray}
\label{b}
\alpha_{i}(\omega) = \alpha^{S}_{i}(\omega) +
 (\frac{3 {\rm cos}^{2} \theta_{p}-1}{2})\frac{3m^{2}_{j_i}-j_i(j_i+1)}{j_i(2j_i-1)} \alpha^{T}_{i}(\omega).
\end{eqnarray}
The dynamic polarizability 
not only depends on the value of $m$ but also the $\theta_{p}$ 
in a certain frequency $\omega$.

In the case of cos$\theta_k=0$, parameters $\theta_{maj}$ and $\theta_{min}$ satisfy the relation
\begin{eqnarray}
\label{q3}
\theta_{\rm{maj}} +\theta_{\rm{min}} =\frac{\pi}{2}.
\end{eqnarray}
With the use of Eq.~(\ref{q4}) and Eq.~(\ref{q3}), Eq.~(\ref{q2}) can be further simplified to
\begin{eqnarray}
\rm{cos}^2\theta_p = \frac 12 + \frac{\sqrt{1-\mathcal{A}^2}}{2}\rm{cos}2\theta_{maj}.
\end{eqnarray}
Therefore, for a given value of $\mathcal{A}$, cos$^2\theta_p$ satisfies
\begin{eqnarray}
\label{range}
\frac 12 - \frac{\sqrt{1-\mathcal{A}^2}}{2} \leq \rm{cos}^2\theta_p\leq  \frac 12 + \frac{\sqrt{1-\mathcal{A}^2}}{2}.
\end{eqnarray}
As seen from Eq.~(\ref{range}), $\mathcal{A}=0$ corresponds to 
$0\leq \rm{cos}^2\theta_p \leq 1$, in which cos$^2\theta_p$ covers 
the largest range [0,1], while |$\mathcal{A}$|=1 just gives rise to 
cos$^2\theta_p =\frac12$.
In the following, we mainly discuss the case of linearly polarized
light with cos$\theta_k=0$ which is numerically equivalent to 
$\mathcal{A}=0$ as seen from Eq.~(\ref{a}). 

Firstly, two particular cases are considered. 
One of them is $\rm{cos}^2\theta_p=1$. This means the $\hat{e}_z$
axis is perpendicular to the wave vector but parallel to the
polarization vector, i.e. $\rm{\hat{e}_z \perp \hat{k}}$ and
$\rm{\hat{e}_z
\parallel \hat{\epsilon}}$. In this case, Eq.(\ref{b}) becomes
\begin{eqnarray}
\label{para}
\alpha_{i}(\omega) = \alpha^{S}_{i}(\omega) +
\frac{3m^{2}_{j_i}-j_i(j_i+1)}{j_i(2j_i-1)} \alpha^{T}_{i}(\omega).
\end{eqnarray}
Another case is $\rm{cos}^2\theta_p=0$, which means the $\hat{e}_z$
axis is perpendicular to the wave vector and the polarization vector, 
i.e. $\rm{\hat{e}_z \perp \hat{k}}$ and $\rm{\hat{e}_z \perp
\hat{\epsilon}}$. 
Similarly, Eq.(\ref{b}) is simplified as follows,
\begin{eqnarray}
\alpha_{i}(\omega) = \alpha^{S}_{i}(\omega) -
\frac{3m^{2}_{j_i}-j_i(j_i+1)}{2j_i(2j_i-1)} \alpha^{T}_{i}(\omega).
\end{eqnarray}
\begin{figure}[tbh]
\vspace{-0.4em}
\label{dynamic}
\setlength{\abovecaptionskip}{0pt}
\setlength{\belowcaptionskip}{0pt}
\centering{
\includegraphics[width=8.5cm]{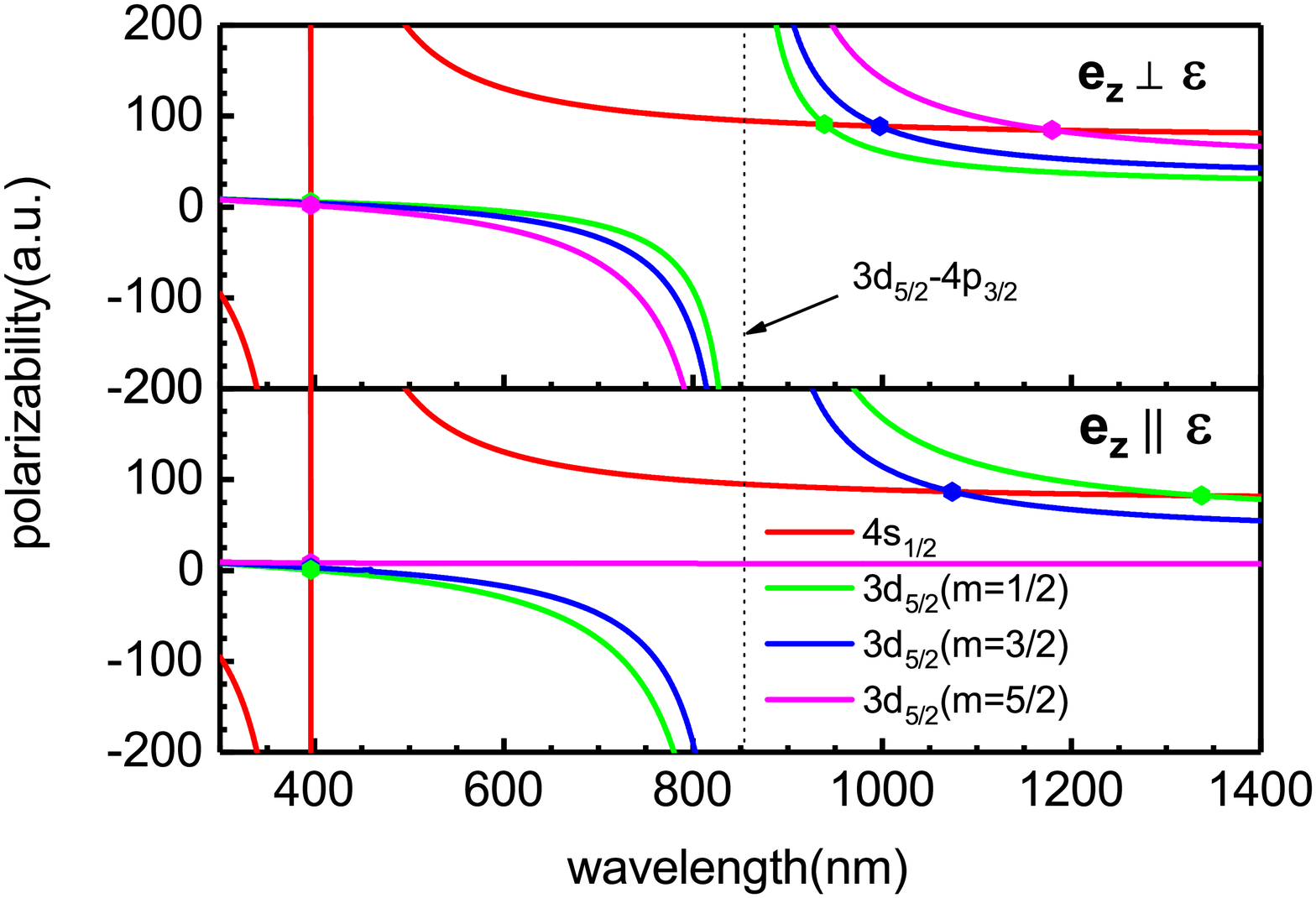}}
\vspace{0.5em}
\caption{Dynamic polarizabilities (in a.u.) of the $4s$ and $3d_{\frac52}(m=\frac12,\frac32,\frac52)$
states in the wavelength range 300 - 1400 nm.
The upper panel plotted the dynamic polarizabilities in the case of
$\hat{e}_z \perp \hat{\epsilon}$.
The lower panel plotted the dynamic polarizabilities 
in the case of $\hat{e}_z \parallel \hat{\epsilon}$.
The approximate position of the $3d_{\frac52}\to4p_{\frac32}$ resonance is
indicated by the vertical dotted line.} 
\end{figure}
\begin{table}
\caption{\label{magic} Magic wavelengths (nm) for the $4s_{\frac12}-3d_{\frac52}$
transitions of the Ca$^{+}$ ions with the linearly polarized light.
$\hat{e}_z \parallel \hat{\epsilon}$ represents the case that
the $\hat{e}_z$ axis is parallel the polarization vector $\hat{\epsilon}$.
$\hat{e}_z \perp \hat{\epsilon}$ represents the case that
the $\hat{e}_z$ axis is perpendicular
to the polarization vector $\hat{\epsilon}$.}
\begin{ruledtabular}
\begin{tabular}{llll}
\multicolumn{3}{c}{ $\hat{e}_z \parallel \hat{\epsilon}$}  & $\hat{e}_z \perp \hat{\epsilon}$    \\
\cline{1-3}  \cline{4-4}
RCICP    & DFCP              & R all- & RCICP   \\
         &\cite{pra-tang13a} &   order\cite{pra-kaur15a}  &\\                                                           
\hline
\multicolumn{4}{c}{ $4s_{\frac12}\to3d_{\frac52}$(m=5/2)}  \\
  &                   &                          & 1179.33(57.92)    \\
395.79485(4)   &395.7968(1)        & 395.79     & 395.79608(9)  \\
\multicolumn{4}{c}{ $4s_{\frac12}\to3d_{\frac52}$(m=3/2)}  \\
1073.80(31.61)& 1074.336(26.352)  & 1052.26    & 997.31(17.41)    \\
395.79584(6)   & 395.7978(1)       & 395.79     & 395.79559(4) \\
\multicolumn{4}{c}{ $4s_{\frac12}\to3d_{\frac52}$(m=1/2)}  \\
1337.31(115.38)& 1338.474(82.593)  & 1271.92    & 938.83(8.97)       \\
395.79633(11)   & 395.7982(1)       & 395.79     & 395.79534(1)  \\
\end{tabular}
\end{ruledtabular}
\end{table}
\begin{figure}[tbh]
\vspace{-0.5cm}
\label{395}
\setlength{\abovecaptionskip}{-4pt}
\setlength{\belowcaptionskip}{-4pt}
\centering{
\includegraphics[width=8.5cm]{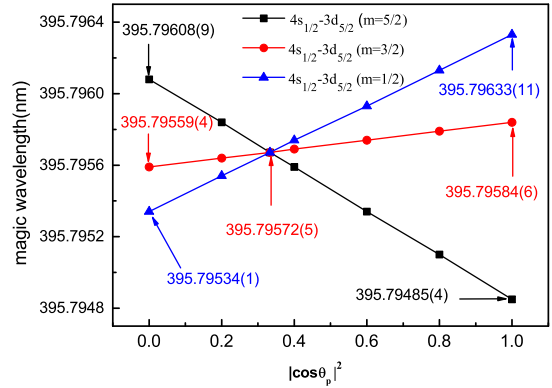}
\vspace{0.5em}}
\caption{ \label{fig1}~The dependence of magic wavelengths, which lie between the
$4s_{\frac12}\to 4p_{\frac12}$ transition wavelength (393.37 nm) and $4s_{\frac12}\to 4p_{\frac32}$ 
transition wavelength (396.85 nm), of each magnetic transition of the $4s_{\frac12}\to 3d_{\frac52}$  
upon cos$^2\theta_p$ in the case of linearly polarized light. }
\end{figure}
\begin{figure}[tbh]
\vspace{-0.2cm}
\label{fig2}
\setlength{\abovecaptionskip}{-4pt}
\setlength{\belowcaptionskip}{-4pt}
\centering{
\includegraphics[width=8.5cm]{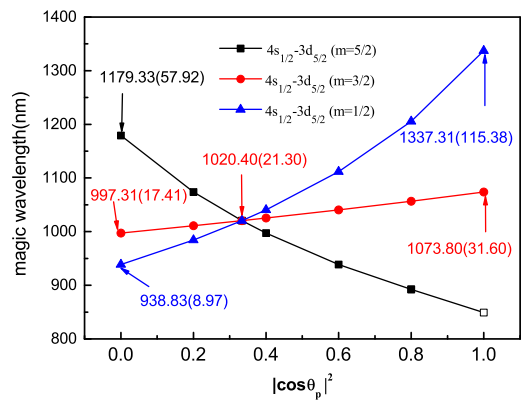}
\vspace{0.5em}}
\caption{ \label{fig1}~Same as Fig~2 but for the 
magic wavelengths longer than the $3d_{\frac52}\to 4p_{\frac32}$ 
transition wavelength (854.21 nm).}
\end{figure}

Figure~2 shows the dynamic polarizabilities of the $4s$ and $3d_{\frac52}$ states 
in the wavelength range 300 - 1400 nm for the laser
polarization direction perpendicular (upper panel) and parallel (lower panel) 
the quantization axis, respectively. 
Since the dynamic polarizability of the $4s_{\frac12}$ state
only has the isotropic scalar part,
the dynamic polarizability of the $4s_{\frac12}$ state in case of $\rm{\hat{e}_z \perp \hat{\epsilon}}$ is
the same as the one in $\rm{\hat{e}_z\parallel \hat{\epsilon}}$. 
However, the dynamic polarizabilities of the
$3d_{\frac52}$ state for both the cases $\rm{\hat{e}_z \perp \hat{\epsilon}}$ and
$\rm{\hat{e}_z\parallel \hat{\epsilon}}$ are completely 
different for each magnetic components 
due to the contribution from the anisotropic tensor part.
For example, 
when the wavelength is close to the $3d_{\frac52}\to 4p_{\frac32}$
resonant transition wavelength (854.21 nm),
the dynamic polarizability of the $3d_{\frac52}(m=\frac52)$ state is infinite 
in the case of  $\rm{\hat{e}_z \perp \hat{\epsilon}}$ but it is
finite in the $\rm{\hat{e}_z \parallel \hat{\epsilon}}$. 
To be more specific, as the explanation in Ref.~\cite{pra-jiang17b}, 
the contributions of the tensor and scalar terms from the $3d_{\frac52}\to 4p_{\frac32}$
transition cancel each other out in the case of 
$\rm{\hat{e}_z \parallel \hat{\epsilon}}$. 
The intersections of the dynamic polarizabilities of the $4s_{\frac12}$ and each magnetic 
states of $3d_{\frac52}$ give rise to magic wavelengths. 
For a given magnetic transition, we can see the magic wavelengths 
are different for
the cases $\rm{\hat{e}_z \perp \hat{\epsilon}}$ and
$\rm{\hat{e}_z\parallel \hat{\epsilon}}$.
Two magic wavelengths have been found for each of $4s \to 3d_{5/2}$ magnetic transitions 
(except for the $4s \to 3d_{5/2} (m=5/2)$ transition with $\rm{\hat{e}_z\parallel \hat{\epsilon}}$, 
which only has one magic wavelength), 
one lies between $4s_{\frac12} \to 4p_{\frac12,\frac32}$ transition wavelengths, 
the other is longer than the $4p_{\frac32}\to 3d_{\frac52}$ transition wavelength (854.21nm).  

Table~I lists the magic wavelengths of the $4s_{\frac12}\to3d_{\frac52}$ transition 
for the two particular cases, $\hat{e}_z
\parallel \hat{\epsilon}$ and $\rm{\hat{e}_z \perp \hat{k}}$, 
along with other available values. For the case of $\rm{\hat{e}_z \perp \hat{k}}$, there
are no existing results for comparison.
It is found that the present results 
by using the RCICP approach are in
good agreement with the results of Tang $et$ $al.$ \cite{pra-tang13a} for the $\hat{e}_z \parallel\hat{\epsilon}$. 
The biggest
difference is just 1.16 nm at 1337.31 nm. 
However, the results of Kaur $et$ $al.$ by using 
relativistic all-order method \cite{pra-kaur15a} have a big
difference from the RCICP and Tang's results. 
For example, the difference of the magic wavelengths 
between the RCICP and relativistc 
all-order results is 65.39 nm at 1337.31 nm, and 21.54 nm at 1073.80 nm. 

Furthermore, we investigate the variation of magic wavelengths 
for the $4s\to 3d_{\frac52}$ transition with cos$^2\theta_p$. Figure~3 shows
the dependence of the magic wavelengths 
near 395.79 nm upon cos$^2\theta_p$, which lie between the $4s_{\frac12}\to 4p_{\frac12}$ and
$4s_{\frac12}\to 4p_{\frac32}$ transition wavelengths.
As shown clearly from Fig~3, the magic wavelengths change nearly linearly with cos$^2\theta_p$.
Also, the difference of magic wavelengths is small for different magnetic transitions. 
Meanwhile, the magic wavelength of each magnetic transition changes
weakly with cos$^2\theta_p$ as well. 
For example, for the $4s_{\frac12}\to3d_{\frac52} (m=5/2)$ transition, 
the difference of magic wavelengths is just 0.0012 nm for cos$^2\theta_p=0$ and cos$^2\theta_p=1$.
The absolute values of derivatives of magic wavelengths for 
the $4s_{\frac12}\to 3d_{\frac52}(m=1/2,3/2,5/2)$, $|\rm{d\lambda_{magic}}/\rm{dcos^2\theta_p|}$,  
are 0.00099, 0.00025, 0.0012, respectively.  
It means these magic wavelengths are not sensitive to the quantity cos$^2\theta_p$.

\begin{table}
\caption{\label{magic3} Magic wavelengths (nm) for the $4s_{\frac12}-3d_{\frac32}$
transitions of the Ca$^{+}$ ions with the linearly polarized light.}
\begin{ruledtabular}
\begin{tabular}{llll}
\multicolumn{3}{c}{ $\hat{e}_z \parallel \hat{\epsilon}$}  & $\hat{e}_z \perp \hat{\epsilon}$    \\
\cline{1-3}  \cline{4-4}
RCICP        & DFCP              & R all- & RCICP   \\
&\cite{pra-tang13a} &   order\cite{pra-kaur15a}  &\\
\multicolumn{4}{c}{ $4s_{\frac12}\to3d_{\frac32}$(m=3/2)}  \\
887.28(3.52) & 887.382(3.196) & 884.54  & 1142.81(44.52)     \\
             &    &  & 851.1325(78)       \\
395.7951(1)  & 395.7970(1)    & 395.79 & 395.79593(7)\\
\multicolumn{4}{c}{ $4s_{\frac12}\to3d_{\frac32}$(m=1/2)}  \\
1307.60(96.2)& 1308.590(71.108) &1252.44 & 956.08(10.23)\\
850.3301(18) & 850.335(2)       & 850.33 & 855.1243(560)\\
395.7962(1)  & 395.7981(1)      & 395.80 & 395.795381(2)\\
\end{tabular}
\end{ruledtabular}
\end{table}
\begin{figure}[tbh]
\vspace{0.5cm}
\label{fig3}
\setlength{\abovecaptionskip}{-4pt}
\setlength{\belowcaptionskip}{-4pt}
\centering{
\includegraphics[width=8.5cm]{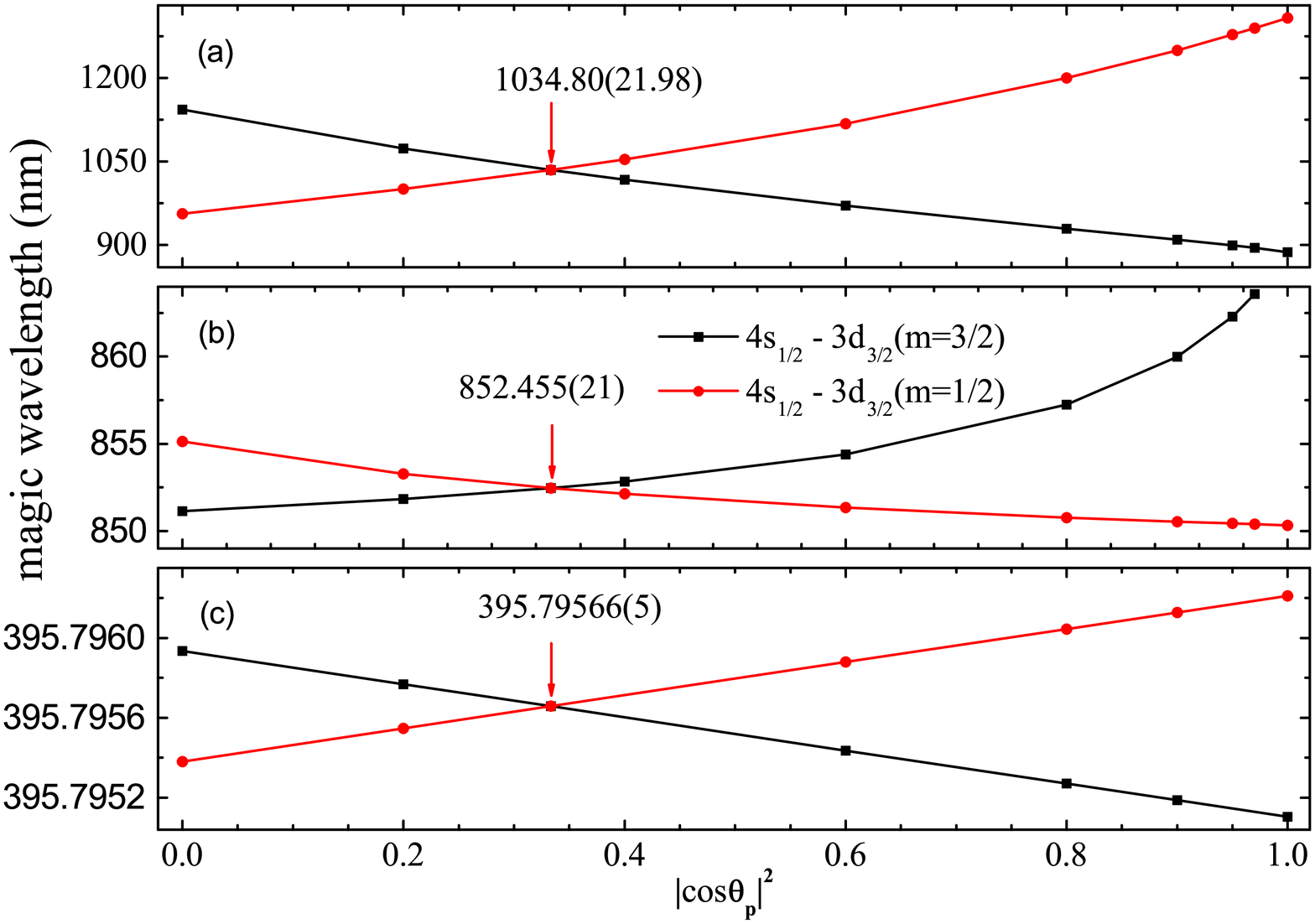}
\vspace{0.5em}}
\caption{ \label{fig1}~The magic wavelengths
of the $4s_{\frac12}\to 3d_{\frac32}$ transition of the
Ca$^{+}$ ions for linearly polarized light.
(a) the magic wavelengths which lie between the
$4s_{\frac12}\to 4p_{\frac12}$ transition wavelength (393.37 nm) and $4s_{\frac12}\to 4p_{\frac32}$
transitions wavelength (396.85 nm);
(b) the magic wavelengths which lie between the
$3d_{\frac32}\to 4p_{\frac32}$ transition wavelength (849.80 nm) and $3d_{\frac32}\to 4p_{\frac12}$ 
transition wavelength (866.21 nm);
(c) the magic wavelengths longer than the $3d_{\frac32}\to 4p_{\frac12}$ 
transition wavelength (866.21 nm). }
\end{figure}

Figure~4 shows the dependence of the other magic wavelengths, which are longer than 
$3d_{\frac32}\to 4p_{\frac32}$ transition wavelength, 
for the $4s_{\frac12}\to3d_{\frac52}$ transition upon cos$^2\theta_p$. 
As seen from Fig~4, the magic wavelengths change nonlinearly. 
The magic wavelengths for the $4s_{\frac12}\to3d_{\frac52}(m=1/2, 3/2)$ transitions 
become longer with the increase of cos$^2\theta_{\rm p}$, 
while the magic wavelengths for the $4s_{\frac12}\to3d_{\frac52}(m=5/2)$
transition become shorter. Moreover, 	
the magic wavelength of each magnetic transition changes
strongly with cos$^2\theta_p$.	
For example, for the $4s_{\frac12}\to3d_{\frac52}(m=1/2)$ transition, 
the difference of magic wavelengths is 398.48 nm for cos$^2\theta_p=0$ and cos$^2\theta_p=1$.
The minimum absolute values of derivatives, $|\rm{d\lambda_{magic}}/ \rm{dcos^2\theta_p|}$,
for the $4s_{\frac12}\to 3d_{\frac52}(m=1/2,3/2,5/2)$ transitions are 228, 77, 217 respectively. 
Thence, these magic wavelength longer than the 
$3d_{\frac52}\to 4p_{\frac32}$ transition wavelength 
varies sensitively with cos$^2\theta_p$.

As shown in Figures~3 and 4, however, different curves intersect at one point. 
The magic wavelength are independent of magnetic sublevels at the intersections. 
Moreover, at the intersections, the contribution of tensor polarizabilities is zero.
This condition can be attained when cos$^2\theta_{p}=\frac13$ 
for a linearly polarized light. 
Under this condition, the angle $\theta_p$ is referred to as "magic angle" \cite{budker} and is given by
\begin{eqnarray}
\theta_p=\rm{arccos}(\frac{1}{\sqrt{3}})\approx 54.74^ \circ.
\end{eqnarray}
According to Eq.~(\ref{range}), the determination of
the "magic angle" requires the condition $|\mathcal{A}| \leq
\frac{2\sqrt{2}}{3}$. 
The magic wavelengths corresponding to the magic angle are determined for the
$4s_{\frac12}\to3d_{\frac52}$ transition as shown in Figures~3 and 4. 
For instance, at the magic angle, the magic wavelength are 395.79572(5) 
and 1024.40(21.30) nm for
the $4s_{\frac12}\to3d_{\frac52}$ transition.

Table~II lists the magic wavelengths for the $4s_{\frac12}\to3d_{\frac32}$
 transition in the cases of $\hat{e}_z \parallel \hat{\epsilon}$ 
and $\hat{e}_z \perp \hat{\epsilon}$. 
Good consistency is obtained for the $\hat{e}_z \parallel \hat{\epsilon}$,
while there are currently no comparable results for $\hat{e}_z \perp \hat{\epsilon}$. 
Figure~5 shows the dependence of the magic
wavelengths for the $4s_{\frac12}\to3d_{\frac32}$ transition upon cos$\rm{^2\theta_p}$.
Similarly, the magic wavelength near 395.79 nm is
insensitive to cos$\rm{^2\theta_p}$, while the 
magic wavelength longer than 854.21 nm strongly depends on cos$\rm{^2\theta_p}$.

\section{Conclusions}

The dynamic polarizabilities of the $4s_{\frac12}$ and $3d_{j}$ 
states  of the Ca$^+$ ions are calculated. The magic wavelengths for the
$4s_{\frac12}\to3d_{\frac32,\frac52}$ transitions are 
identified for $\hat{e}_z \perp \hat{\epsilon}$ and $\hat{e}_z \parallel
\hat{\epsilon}$ in the case of a linearly polarized light (cos$\rm{\theta_k}=0$ or $\mathcal{A}=0$).
The dependence of the magic wavelengths upon the cos$\rm{^2\theta_p}$
are analyzed. 
It is found that the magic wavelength near
395.79 nm is insensitive to the angle between the quantization
axis $\hat{e}_z$ and the polarization vector $\hat{\mathbf{\epsilon}}$. 
In contrast, the magic wavelength longer than the $3d_{\frac52}\to4p_{\frac32}$ transition wavelength (854.21 nm) is very sensitive to
cos$\rm{^2\theta_p}$ due to the contribution 
of the anisotropic tensor polarizability. 
Meanwhile, we find that some particular magic wavelengths are
independent of magnetic sublevels at the magic angle.

\section{Acknowledgments}
This work was supported by the National Key Research and Development Program of China under 
Grant No.2017YFA0402300 and National
Natural Science Foundation of China (NSFC) (Grant Nos.11564036, 11774292, U1530142, 
11464042) and 
the Young Teachers Scientific Research Ability Promotion Plan of Northwest Normal University (NWNU-LKQN-15-3).


\end{document}